\documentclass[twocolumn,floatfix,secnumarabic,amssymb,nobibnotes,aps,prd]{revtex4-2}
\usepackage{amsmath,amssymb}
\usepackage{graphicx}
\usepackage{bm}
\usepackage{color}
\usepackage{mathrsfs}
\usepackage{comment}
\usepackage{float}
\usepackage{tabularx}
\usepackage{ulem}
\usepackage[all]{xy}
\usepackage{tikz-cd}
\usepackage{braket}
\usepackage{subcaption}
\usepackage{overpic}
\usetikzlibrary{positioning,arrows.meta,decorations.pathreplacing,calc, ,fit,backgrounds}
\AtBeginDocument{\colorlet{BLUE}{blue}}

\begin{document}

\title{Velocity gradient tensor reconstruction from tracer orientations via Jeffery dynamics}
\author{Isshin Arai}%
\email{k078403@kansai-u.ac.jp}
\affiliation{1 Graduate School of Science and Engineering, Kansai University, Osaka, 564-8680, Japan}
\author{Tomoaki Itano}%
\affiliation{2 Department of Pure and Applied Physics, Faculty of Engineering Science, Kansai University, Osaka, 564-8680, Japan}
\date{\today}

\begin{abstract}
The velocity gradient tensor is a local quantity, but methods that obtain it by spatial differencing of the velocities of spherical tracers amplify measurement noise.
We propose a method that reconstructs the velocity gradient tensor from the orientations of non-spherical tracers such as flakes and rods. Based on the Jeffery-type equation, we show that, for incompressible flow, all components of the velocity gradient tensor are uniquely determined from the orientations of four tracers and their time derivatives in the neighbourhood of a reference point (orientation-based method). The method uses neither tracer positions nor spatial differencing. It therefore requires orientation measurements only near the reference point, without requiring measurements throughout a three-dimensional region, and measurement noise is amplified only through temporal differencing. Consequently, the minimum achievable error scales as $\sigma^{1/2}$ with the noise level $\sigma$, compared with $\sigma^{1/3}$ for the positional PTV-plus-differencing approach (position-based method). Numerical tests further show that the vorticity is reconstructed more accurately than by the position-based method, whereas the strain rate is reconstructed with comparable accuracy. Finally, applying the method to spherical Couette flow, we find that the reconstruction fails where the neighbouring tracers align, which renders the inversion ill-conditioned, and that this degeneracy appears as the characteristic light-and-dark pattern familiar from flow visualization.
\end{abstract}

\maketitle

\section{Introduction}

Tracer particles are widely used to visualize flow fields \citep{Dy82}.
Particle image velocimetry (PIV) and particle tracking velocimetry (PTV) reconstruct the velocity field from the positions of spherical tracers.
Non-spherical tracers, in particular flakes or rods, also encode flow information in their orientation.
The orientation dynamics of non-spherical particles in turbulence have been increasingly well understood~\cite{Puj17, Cui26}, and particle orientations have been used to measure statistical properties of the flow~\cite{Bri21}.
A classical visualization technique records the bias of flake orientations as a two-dimensional light-and-dark pattern \citep{Sa85, Egb95, Gau98, Got11, Yos23, Arai24}.
Such patterns reflect the spatial structure of the flow and are valuable for qualitative interpretation, such as identifying flow transitions, but their quantitative exploitation has remained limited.
For such a tracer sufficiently small compared with the characteristic scale of the flow, the translational motion follows the Lagrangian trajectory of fluid elements, while the rotational motion is governed by a Jeffery-type equation \citep{Je1922, Br62, Le15}.
The orientation is thus driven by the velocity gradient tensor, and is intimately connected to the deformation of material line and surface elements \citep{Bat67, Gi90, Dr91}.
Thus, tracer orientations carry direct information about the velocity gradient.

Extracting this information, however, is obstructed by a non-uniqueness.
Even when random factors, such as thermal fluctuations, tracer-size and -shape variability and measurement noise, are neglected, an orientation of a tracer at a given instant is not determined solely by the velocity gradient at that instant. Under the same velocity-gradient field, different initial orientations or histories generate different trajectories in orientation space, whose dynamics admits a variety of behaviours, including steady, circular, periodic and quasi-periodic attractors \citep{Sze93p, Sze93, Sze94, Arai25, Arai26}.
When random forces such as thermal noise are additionally taken into account \citep{Kida14, Ita25}, the dependence on the initial orientation weakens in time but the orientation becomes a probabilistic distribution.
A measured orientation and its time evolution do not uniquely determine the velocity gradient.
This absence of a one-to-one correspondence---an issue of identifiability \citep{Mac19, Wi21} and reconstructability \citep{Arai26g, Arai_arxiv}---has been the principal obstacle to quantitative, orientation-based flow measurement.
Resolving it is worthwhile because orientations provide direct access to the velocity gradient tensor, a local flow quantity. By contrast, methods that reconstruct the velocity gradient through spatial differencing of the velocity field amplify the errors incurred in the velocity evaluation.
This difficulty has motivated considerable recent effort:
volumetric measurements resolving all nine components \citep{Kru14},
gradients defined from a tetrad of tracked points \citep{Yan20},
regression of gradients directly from particle trajectories \citep{Har24},
and, beyond spherical tracers, rods and fibers whose motion is coupled
to the local velocity gradient \citep{Ni15, Cav20, Bri21}.
Ni et al.~\cite{Ni15} simultaneously measured the tumbling of rods and the surrounding velocity gradient tensor in turbulence and examined their relationship.
Cavaiola et al.~\cite{Cav20} numerically showed that the full velocity gradient tensor can be reconstructed from the end-to-end velocity differences of an assembly of rigid fibers.
Brizzolara et al. \cite{Bri21} measured two-point statistics of turbulence accurately using the velocity difference between the two ends of rigid fibers, which is unaffected by the relative dispersion of particles.

In this paper, we present a method that determines all components of the velocity gradient tensor in incompressible flow from the orientations of axisymmetric particles, by inverting Jeffery’s equation.
Exploiting the fact that an orientation is a three-dimensional vector, the full velocity gradient tensor in incompressible flow is reconstructed from a low-dimensional measurement region without positional information.
Specifically, we show that all its components can be obtained uniquely from four orientations and their time derivatives at a single point, provided that the four orientations are sufficiently distinct in direction.
When they instead degenerate towards a specific direction, the reconstruction becomes ill-conditioned and fails.

Since the required orientations cannot be sampled exactly at a single point but must be supplied by neighbouring particles, we analyse the error introduced by this substitution and compare the method with the conventional approach based on spatial differencing of particle positions.
The comparison is carried out both theoretically and through numerical simulations of the ABC flow~\citep{Dom86}.
In the presence of measurement noise, the position-based method amplifies the noise because the velocity differences are divided by the particle separation, whereas the present method avoids this amplification.
Consequently, the minimum attainable error scales with the noise intensity according to different power laws in the two methods.
Furthermore, under conditions where the two methods yield comparable errors in the full velocity gradient tensor, only the present method exhibits a clear separation between the vorticity and strain contributions, which originates from the Jeffery orientation dynamics: the vorticity is reconstructed more accurately than the strain.
Finally, as an application to a real flow, we reconstruct the velocity gradient tensor in the spherical Couette flow (SCF)~\cite{Egb95}, and demonstrate concretely how particle orientations degenerate as they drift through the flow, rendering the reconstruction increasingly difficult.
Interestingly, this behaviour reveals a direct correspondence with the classical light-and-dark visualization discussed above. The formation of a characteristic light-and-dark pattern is itself a signature of orientational degeneracy.
Consequently, wherever such a pattern is visible the velocity gradient cannot be reconstructed, whereas wherever no pattern is discernible the reconstruction succeeds.\\

The remainder of this paper is organized as follows.
In section~\ref{section:method}, based on the Jeffery-type equation, we present the formulation that reconstructs the velocity gradient tensor uniquely from four orientations and their time derivatives.
In section~\ref{section:error}, we show how truncation and measurement noises propagate to the reconstructed velocity gradient in the present method and in the positional PTV-plus-differencing approach.
In section~\ref{section:results}, we examine through numerical simulations how the reconstruction error depends on the measurement noise and on the degeneracy of the orientations.
Conclusions and outlook are given in section~\ref{section:conclusion}.

\section{Method}\label{section:method}

Without thermal fluctuations and interactions between particles, 
for an infinitesimally thin flake-like particle with a sufficiently small Stokes number --- defined as the ratios of the particle response time to the characteristic flow time scale for both translation and rotation --- rather than one, the governing equations for its translational and rotational motion are expressed as
Neglecting thermal fluctuations and interactions between particles, for an infinitesimally thin flake-like particle with Stokes numbers much smaller than one --- defined as the ratios of the particle response times to the characteristic flow time scale for translation and rotation --- the governing equations for its translational and rotational motion are expressed as
\begin{equation}
\dot {\bm r} = \bm {u}(\bm r),\label{eq:position}
\end{equation}
\begin{equation}
\dot {\bm s} = \bm s \times \bm s \times(\nabla \bm u(\bm r) \cdot \bm s).\label{eq:orientation}
\end{equation}

Here, $\bm {r}$ is the particle position, and $\bm {u}(\bm r)$ and $\nabla \bm u(\bm r)$ are the flow velocity and the velocity gradient tensor at that position, respectively.
The vector $\bm s$ is the unit normal vector of the flakes, and is driven by the velocity gradient tensor as in Eq.~\eqref{eq:orientation} \citep{Got11}.
Note that Eq.~\eqref{eq:orientation} corresponds to the Jeffery-type equation~\citep{Je1922, Br62, Le15} for an infinitely thin disk. For axisymmetric particles with a finite aspect ratio, the equation takes a more general form containing a shape factor.
In this paper we treat this limiting form.
For a rod, whose orientation vector $\bm s_{\mathrm{rod}}$ is taken along the particle axis, its rotational motion is $\dot{\bm s}_{\mathrm{rod}}=-\bm s_{\mathrm{rod}}\times \bm s_{\mathrm{rod}}\times(\nabla \bm u^T\cdot \bm s_{\mathrm{rod}})$.
Therefore, the following reconstruction method can also be applied to rods.
Although rods are less reflective than flakes, their orientations are often easier to measure accurately.

Equation~\eqref{eq:orientation} is rewritten as
\begin{equation}
G\cdot \bm{s}^{(i)}  -\lambda^{(i)}\, \bm{s}^{(i)} = -\dot{\bm{s}}^{(i)}.
\label{eq:orientations}
\end{equation}
Here $G:=\nabla \bm u$, and the superscript $i$ labels the time-evolution equations for different orientations on a unit sphere.
Furthermore, $\lambda^{(i)}:=\bm s^{(i)}\cdot G \cdot \bm s^{(i)}$.
Although $\lambda^{(i)}$ is a quantity determined by $G$, treating it as an independent unknown allows the following system to be formulated as a linear problem for $G$.

Each tracer provides three equations for the nine components of \(G\), along with one additional unknown \(\lambda^{(i)}\).
With \(n\) tracers, there are \(3n+1\) equations, including the trace-free condition, for \(9+n\) unknowns. Thus, the numbers of equations and unknowns are equal when \(n=4\).
The additional equation is the trace-free condition,
\(\mathrm{tr}(G)=G_{xx}+G_{yy}+G_{zz}=0,\)
i.e.\ incompressibility, \(\nabla\cdot\bm{u}=0\).
Thus, four pairs \(\{\bm{s}^{(i)},\dot{\bm{s}}^{(i)}\}\), with \(i=1,2,3,4\), provide \(3\times4=12\) equations.

We define the unknown vector $\mathbf{x}\in\mathbb{R}^{13}$ as
\begin{equation}
\begin{split}
\mathbf{x} := [\,&G_{xx}, G_{xy}, G_{xz}, G_{yx}, G_{yy}, G_{yz}, G_{zx}, G_{zy}, G_{zz},\\
&\lambda^{(1)}, \lambda^{(2)}, \lambda^{(3)}, \lambda^{(4)}\,]^T,
\end{split}
\label{eq:x}
\end{equation}
and the vector $\mathbf{q}\in\mathbb{R}^{13}$ as
\begin{equation}
\mathbf{q} := [-\dot{\bm s}_x, -\dot{\bm s}_y, -\dot{\bm s}_z, 0]^T,\label{eq:q}
\end{equation}
where $\dot{\bm s}_j:=(\dot{s}^{(1)}_j,\dot{s}^{(2)}_j,\dot{s}^{(3)}_j,\dot{s}^{(4)}_j)$ for $j=x,y,z$. The $13\times13$ coefficient matrix $\mathbf{M}$ is
\begin{equation}
\mathbf{M} := \left(\begin{array}{ccc|c}
\mathbf{S} & \mathbf{0} & \mathbf{0} & \mathbf{D}_x \\
\mathbf{0} & \mathbf{S} & \mathbf{0} & \mathbf{D}_y \\
\mathbf{0} & \mathbf{0} & \mathbf{S} & \mathbf{D}_z \\
\hline
\mathbf{e}_x & \mathbf{e}_y & \mathbf{e}_z & 0
\end{array}\right),\label{eq:B}
\end{equation}
where
\begin{align*}
\mathbf{S}
&:=
\left(
\begin{array}{ccc}
s^{(1)}_x & s^{(1)}_y & s^{(1)}_z \\
s^{(2)}_x & s^{(2)}_y & s^{(2)}_z \\
s^{(3)}_x & s^{(3)}_y & s^{(3)}_z \\
s^{(4)}_x & s^{(4)}_y & s^{(4)}_z
\end{array}
\right),\\
\mathbf{D}_j &:= \operatorname{diag}(-s^{(1)}_j,-s^{(2)}_j,-s^{(3)}_j,-s^{(4)}_j)\quad(j=x,y,z),\\
\mathbf{e}_x &:= (1,0,0),\;
\mathbf{e}_y :=(0,1,0),\;
\mathbf{e}_z :=(0,0,1).
\end{align*}
The linear system $\mathbf{M}\mathbf{x}=\mathbf{q}$ has, when $\mathbf{M}$ is non-singular, the unique solution
\begin{equation}
\mathbf{x} = \mathbf{M}^{-1}\mathbf{q},\label{eq:reconstruction}
\end{equation}
determining the nine components of $G$ and $\lambda^{(1)},\dots,\lambda^{(4)}$. The condition $\det(\mathbf{M})\neq0$ is precisely the condition for unique determination. It requires that no three of the four orientations $\bm{s}^{(1)},\dots,\bm{s}^{(4)}$ be coplanar (any three span three-dimensional space). Four orientations along the vertex directions of a regular tetrahedron, for example, satisfy this condition.

\section{Comparison of the noise propagation between the positional and orientational methods}
\label{section:error}
In practice, the time derivatives and spatial gradients can be estimated by least-squares fitting using data from multiple tracers and time instants.
Here, to compare the two methods, we use the same minimal dataset: four tracers observed at two instants, $t$ and $t+\Delta t$ (see Appendices~\ref{section:error_position} and~\ref{section:error_orientation}).

\subsection{Tracking positions}
First, consider evaluating the velocity gradient tensor from the particle positions via the velocity field (hereafter called the position-based method).
In this method, the velocity is obtained from the particle positions by temporal differencing, and the velocity gradient is then obtained by spatial differencing of the velocity (see Appendix~\ref{section:error_position} for a detailed analysis).

The measured position is assumed to contain position noise $\bm\epsilon_r$, $\tilde{\bm r}:=\bm r+\bm\epsilon_r$, from which the velocity is reconstructed by the forward difference
\[
\tilde{\bm u}
:=
\frac{\tilde{\bm r}(t+\Delta t)-\tilde{\bm r}(t)}{\Delta t}
\]
with the time interval $\Delta t$.
The reconstructed velocity gradient tensor $\tilde{G}$ is obtained by further spatial differencing of the reconstructed velocity as
\[
\tilde G
=
\frac{
\tilde{\bm u}(\bm r+\Delta\bm r)
-
\tilde{\bm u}(\bm r)
}
{\Delta\bm r},
\]
where $\Delta\bm r$ is the spatial interval and the right-hand side is a shorthand for componentwise forward differences.

To leading order in the small quantities, the error of the reconstructed velocity gradient $\tilde{G}$ relative to its true value $G$ is
\begin{align}
\delta G:={}&\tilde G-G\notag\\
\approx{}&
O\biggl(
\Bigl(
\frac{\partial G}{\partial t}
+
G^2
\Bigr)\Delta t
+
(\Delta\bm l\cdot\nabla)G
\biggr)\notag\\
&+O\bigl(
(\Delta\bm r\cdot\nabla)G
\bigr)
+
O\!\left(
\frac{\|\bm\epsilon_r\|}
{\|\Delta\bm r\|\,\Delta t}
\right),\label{eq:PTVerror}
\end{align}
where $\Delta\bm l=\bm u\,\Delta t+O(\Delta t^2)$ is the particle displacement over the time interval $\Delta t$.
The order symbol $O(\cdot)$ is applied to each component of a tensor.
The first and second terms on the right-hand side are the truncation errors due to the temporal and spatial differencing, respectively, while the third term shows that the noise $\bm\epsilon_r$ is amplified in the form of division by the product of the spatial interval $\|\Delta\bm r\|$ and the time interval $\Delta t$.  Here, $\|\cdot\|$ denotes the Euclidean norm for vectors.
Because of the forward difference, the truncation error in the velocity
is proportional to the particle acceleration, and the first term
on the right-hand side arises when this error is further differenced in space;
the term $(\Delta\bm l\cdot\nabla)G$ originates from the advective part of the acceleration.

\subsection{Tracking orientations}\label{subsec:tracking-orientation}
The orientation method proposed in Eq.~\eqref{eq:reconstruction} uses the matrix $\mathbf{M}(\bm s)$ constructed from the particle orientations and the vector $\mathbf{q}(\dot{\bm s})$ constructed from their time derivatives (hereafter called the orientation-based method).
Unlike the position-based method, it does not involve spatial differencing of the velocity field, so the error propagation path also differs (see Appendix~\ref{section:error_orientation} for a detailed analysis).

At $\tilde{\bm r}=\bm r+\Delta \bm r$, the measured orientation is assumed to contain orientation noise $\bm\epsilon_s$, $\tilde{\bm s}:=\bm s+\bm\epsilon_s$, from which the time derivative is reconstructed by the forward difference
\[
\tilde{\dot{\bm s}}
:=
\frac{\tilde{\bm s}(t+\Delta t)-\tilde{\bm s}(t)}{\Delta t}.
\]
We also define
\[
\dot{\tilde{\bm s}}
:=
\bm f(\tilde{\bm s};G(\bm r)),
\]
where $\bm f$ denotes the right-hand side of Eq.~\eqref{eq:orientation}; that is, $\dot{\tilde{\bm s}}$ is the value obtained from
Eq.~\eqref{eq:orientation} for the measured orientation
$\tilde{\bm s}$ with the true velocity gradient.
Note that $G$ can be correctly reconstructed as long as
$\dot{\tilde{\bm s}}$ corresponding to $\tilde{\bm s}$ is available, even though $\tilde{\mathbf M}$ is constructed from the noisy orientations.
To leading order in the small quantities, the norm of the error $\delta \dot{\bm s}:=
\tilde{\dot{\bm s}}-\dot{\tilde{\bm s}}$
is
\begin{align}
\|\delta\dot{\bm s}\|
\approx{}&
O\!\left(
\Bigl(
\Bigl\|\frac{\partial G}{\partial t}\Bigr\|
+
\|G\|^2
\Bigr)
\Delta t
+
\|\nabla G\|\,\|\Delta\bm l\|
\right)
\notag\\
&+
O\bigl(\|\nabla G\|\,\|\Delta\bm r\|\bigr)
+
O\!\left(\frac{\|\bm\epsilon_s\|}{\Delta t}\right)
+
O\bigl(\|G\|\,\|\bm\epsilon_s\|\bigr). \label{eq:Flakeerror}
\end{align}
The norm $\|\cdot\|$ denotes the Euclidean norm for vectors and the Frobenius norm for tensors; for the third-order tensor $\nabla G$, it is defined as $\|\nabla G\| := \bigl(\sum_{i,j,k} (\partial_k G_{ij})^2\bigr)^{1/2}$.
The first term is the truncation error of the temporal differencing used to evaluate $\tilde{\dot{\bm s}}$, which is of $O(\Delta t)$.
The second term arises because the orientation at $\bm r$ is substituted by that of a particle located at a spatially different position $\bm r+\Delta\bm r$, and is of $O(\|\Delta\bm r\|)$.
The position noise $\bm\epsilon_r$ is not amplified in the orientation-based method and only shifts the sampling point, contributing $O(\|\Delta\bm r\|)$ for $\|\bm\epsilon_r\|\lesssim\|\Delta\bm r\|$.
The third term shows that the noise $\bm\epsilon_s$ is amplified only through division by the time interval $\Delta t$; unlike the position-based method in Eq.~\eqref{eq:PTVerror}, it is not further divided by the spatial interval $\|\Delta\bm r\|$, so the measurement noise is not amplified by the spatial substitution.
The term $O(\|G\| \|\bm\epsilon_s\|)$, which arises because $\dot{\tilde{\bm s}}$ is evaluated at the noisy orientation, is omitted in Eq.~\eqref{eq:Flakeerror}, since it is smaller than the third term by a factor of $\|G\|\,\Delta t\ll1$.

In deriving Eq.~\eqref{eq:Flakeerror}, for a particle located at $\tilde{\bm r}$ with the measured orientation $\tilde{\bm s}$, the error is decomposed as
\begin{align*}
\delta\dot{\bm s}&=\tilde{\dot{\bm s}}
-\bm f\bigl(\tilde{\bm s};G(\bm r)\bigr)
\\
&=
\Bigl[
\tilde{\dot{\bm s}}
-\bm f\bigl(\tilde{\bm s};G(\tilde{\bm r})\bigr)
\Bigr]
+
\Bigl[
\bm f\bigl(\tilde{\bm s};G(\tilde{\bm r})\bigr)
-\bm f\bigl(\tilde{\bm s};G(\bm r)\bigr)
\Bigr],
\end{align*}
where the first bracket collects the temporal truncation error, $O(\Delta t)$, and the measurement noise, $O(\|\bm\epsilon_s\|/\Delta t)$, at the position $\tilde{\bm r}$, corresponding to the first and third terms of Eq.~\eqref{eq:Flakeerror}, and the second bracket is the spatial substitution error, $O(\|\nabla G\|\,\|\Delta\bm r\|)$, appearing as the second term.

Since the velocity gradient tensor is reconstructed by Eq.~\eqref{eq:reconstruction}, the error $\delta\dot{\bm s}$ of the right-hand-side vector in Eq.~\eqref{eq:q} propagates through $\tilde{\mathbf M}^{-1}$, where $\tilde{\mathbf M}$ is the matrix constructed from the measured orientations $\tilde{\bm s}$, giving
\[
\|\delta G\|
\lesssim
\kappa\,\|\delta\dot{\bm s}\|,
\]
where $\kappa$ is the largest singular value of $\tilde{\mathbf M}^{-1}$.
It depends on the geometric configuration of the four orientations and grows as they approach degeneracy.

Comparing Eqs.~\eqref{eq:PTVerror} and~\eqref{eq:Flakeerror} clarifies the essential distinction between the two methods.
When the measurement noise is present, the difference becomes decisive: the noise term is divided by $\|\Delta\bm r\|\,\Delta t$ in the position-based method, but by $\Delta t$ alone in the orientation-based method.
The spatial interval $\Delta\bm r$ thus amplifies the measurement noise only in the position-based method, which is the essential advantage of the orientation-based method.

\section{Numerical simulation}\label{section:results}

\subsection{Noise propagation and scaling}\label{subsec:ABC}
We examine the order estimates of the error and scaling using the Arnold--Beltrami--Childress (ABC) flow, a steady three-dimensional flow that nonetheless produces diverse particle dynamics depending on the parameters $A,B,C$ \citep{Dom86}:
\begin{align*}
u_x &= A\sin z + C\cos y,\\
u_y &= B\sin x + A\cos z,\\
u_z &= C\sin y + B\cos x.
\end{align*}

In the simulations we take $(A,B,C)=(1,\sqrt{2/3},\sqrt{1/3})$ and reconstruct the velocity gradient tensor. The initial position $r_0=(x,y,z)$ was varied over a uniform $10\times10\times10$ grid of $10^3$ points in the domain $[0,2\pi)^3$. For both methods we place a reference point and three points offset by $\Delta r$ along the $x$-, $y$- and $z$-axes. The position-based method reconstructs the gradient from these four positions by forward differences, while the orientation-based method assigns to each point one of four orientations pointing to the vertices of a regular tetrahedron, ensuring the non-singularity of $\mathbf{M}$ ($\det\mathbf{M}\neq0$, with $\kappa\approx2.19$). In practice, however, the orientation configuration depends on the particle trajectory. In general flows, orientations that are initially distributed isotropically do not necessarily converge to a particular orientation as they drift along the same trajectory. On the other hand, we consider ideal configurations without scatter for both methods. The aim of this study is not to replicate precise experimental conditions, but to clarify the fundamental differences between the two methods.

Equations~\eqref{eq:position} and \eqref{eq:orientation} were integrated by the fourth-order Runge--Kutta method with $dt=2^{-14}$
for all combinations of $\Delta t/dt\in\{1,2,5\}\times10^{m}\,(m=0,1,2)$ and $\Delta r\in\{1,2,5\}\times10^{-m}\,(m=1,2,3,4)$. Measurement noise is added as isotropic Gaussian noise, independently at each time and observation point,
\[
\tilde{\bm r}=\bm r+\bm\xi,\ \ \bm\xi\sim\mathcal{N}(\bm 0,\sigma_r^2 I_3),
\quad
\tilde{\bm s}=\frac{\bm s+\bm\eta}{\|\bm s+\bm\eta\|},\ \ \bm\eta\sim\mathcal{N}(\bm 0,\sigma_s^2 I_3),
\]
where $\sigma_r$ and $\sigma_s$ are the standard deviations of the position and orientation noise,
after which $\tilde{\bm u}=(\tilde{\bm r}(t+\Delta t)-\tilde{\bm r}(t))/\Delta t$ and $\tilde{\dot{\bm s}}=(\tilde{\bm s}(t+\Delta t)-\tilde{\bm s}(t))/\Delta t$ are formed by forward differences. Note that $\sigma_r$ and $\sigma_s$ are physically distinct and not directly comparable.

We define the error common to both methods, which arises without measurement noise:
\[
  \varepsilon_0 := \frac{1}{\|G\|}\left[
  \left(\left\|\frac{\partial G}{\partial t}\right\|+\|G\|^2\right)\Delta t
  + \|\nabla G\|\,\bigl(\overline{\Delta l} + \Delta r\bigr)
  \right].
\]
Here, $\Delta t$ is the time interval, $\overline{\Delta l}$ is the ensemble-averaged particle displacement during $\Delta t$, and $\Delta r$ is the spatial separation used for the finite difference.
Measurement noise adds a method-dependent term to $\varepsilon_0$.
We thus define the error estimates of the two methods as
\begin{align}
  \varepsilon_{\mathrm{pos}}
  &:= \varepsilon_0 + \frac{1}{\|G\|}\,\frac{\sigma_r}{\Delta r\,\Delta t},
  \label{eq:epsptv}\\
  \varepsilon_{\mathrm{ori}}
  &:= \varepsilon_0 + \frac{1}{\|G\|}\,\frac{\sigma_s}{\Delta t}.
  \label{eq:epsflake}
\end{align}
The two methods differ only in the noise terms, which originate from Eqs.~\eqref{eq:PTVerror} and \eqref{eq:Flakeerror}.

The relative error is defined as
\begin{equation}
\varepsilon:=\frac{\|\tilde G-G\|}{\|G\|},\label{eq:eps}
\end{equation}
where $\tilde G$ is the reconstructed velocity gradient tensor.\\
\begin{figure*}[t]
\centering
\setlength{\abovecaptionskip}{4pt}
\begin{subfigure}[t]{0.49\linewidth}
\centering
\begin{overpic}[width=\linewidth]{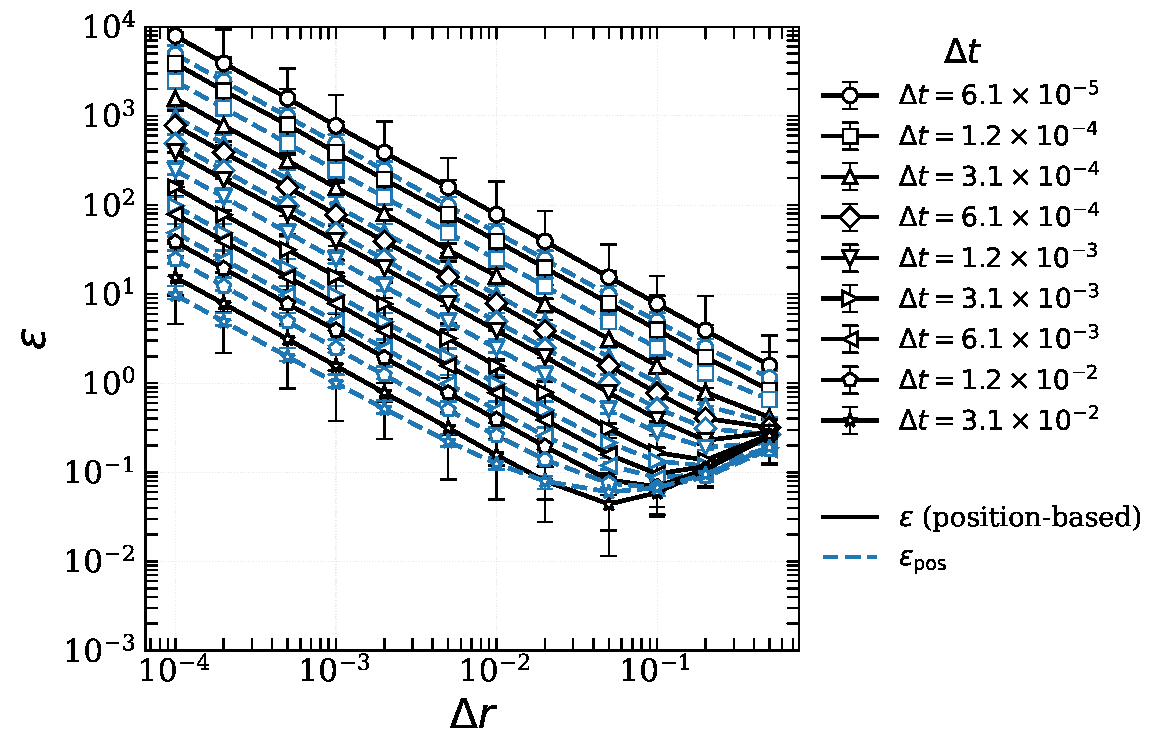}
\put(-5,60){($a$)}
\end{overpic}
\vspace{-\baselineskip}
\phantomcaption\label{fig:PTVnoise}
\end{subfigure}\hfill
\begin{subfigure}[t]{0.49\linewidth}
\centering
\begin{overpic}[width=\linewidth]{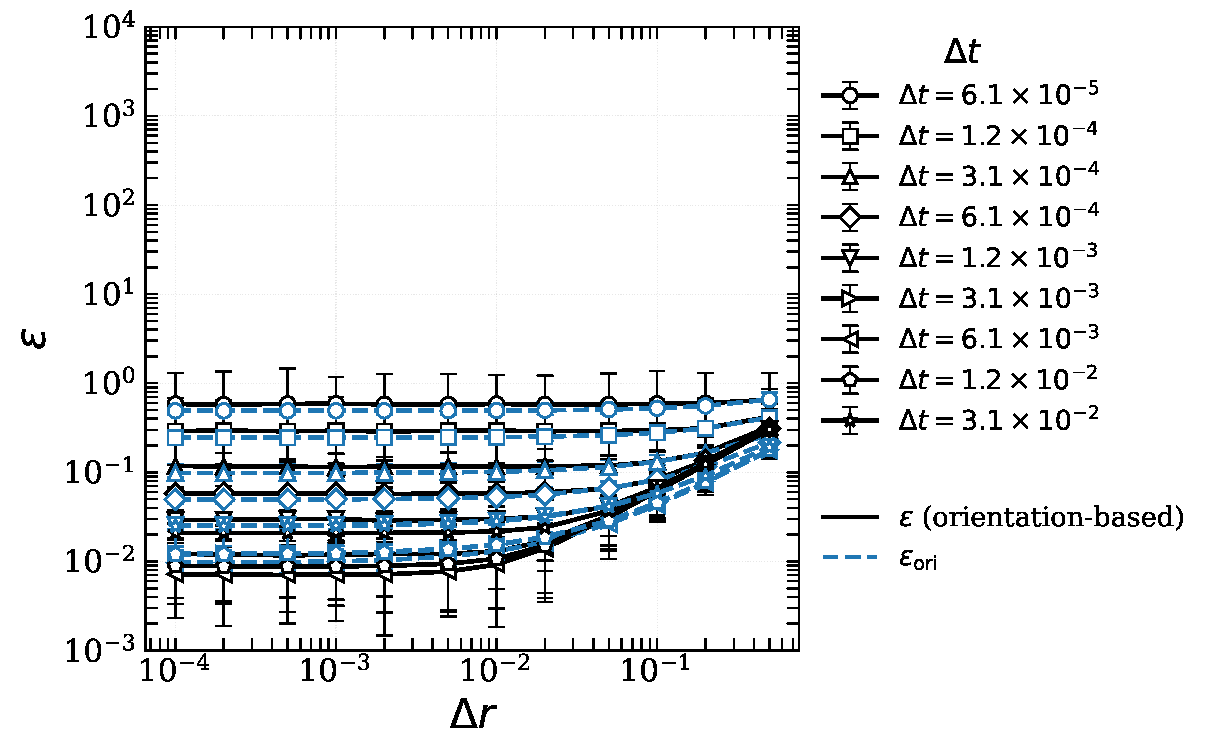}
\put(-5,60){($b$)}
\end{overpic}
\vspace{-\baselineskip}
\phantomcaption\label{fig:Flakenoise}
\end{subfigure}
\caption{Relative error $\varepsilon$ with measurement noise ($\sigma_r=\sigma_s=10^{-5}$). Horizontal axis: spatial interval $\Delta r$; each curve is a time interval $\Delta t$. Solid: relative error, Eq.~\eqref{eq:eps}; dashed: predictions by equations~\eqref{eq:epsptv} and \eqref{eq:epsflake}. ($a$) In the position-based method the noise is amplified as $1/\Delta r$, so $\varepsilon$ rises sharply at small $\Delta r$ and a minimum-error valley appears at intermediate $\Delta r$ through balance with the truncation error. ($b$) In the orientation-based method the noise is independent of $\Delta r$, so no valley appears and the error is not amplified down to fine $\Delta r$.}
\label{fig:epsilon_noise5}
\end{figure*}

Figure~\ref{fig:epsilon_noise5} shows the relative error $\varepsilon$ as a function
of the spatial interval $\Delta r$ for several time intervals $\Delta t$,
with measurement noise $\sigma_r=\sigma_s=10^{-5}$. The error bars indicate the minimum
and maximum values over the ensemble of $10^3$ reference points. The relative errors $\varepsilon$ (solid lines)
agree well with the predictions of  $\varepsilon_{\mathrm{pos}}$ and
$\varepsilon_{\mathrm{ori}}$ (dashed lines).

For the position-based method (Fig.~\ref{fig:epsilon_noise5}(a)), the velocity gradient
is obtained by differencing the velocities of particle pairs separated by
$\Delta r$, so the measurement noise is amplified as $1/\Delta r$. As a result,
$\varepsilon$ increases sharply as $\Delta r$ decreases. At large $\Delta r$,
in contrast, the truncation error of the finite difference becomes dominant.
The balance between these two contributions produces a minimum-error valley
at intermediate $\Delta r$, which implies that $\Delta r$ and $\Delta t$ must
be chosen carefully to achieve the best accuracy.
For the orientation-based method (Fig.~\ref{fig:epsilon_noise5}(b)), the measurement noise is not amplified at small $\Delta r$, no valley appears, and the error decreases monotonically toward the level governed by the time interval $\Delta t$.
This indicates that the orientation-based method is robust to the choice of the spatial interval.\\

\begin{figure}[h]
\centering
\includegraphics[width=0.8\linewidth]{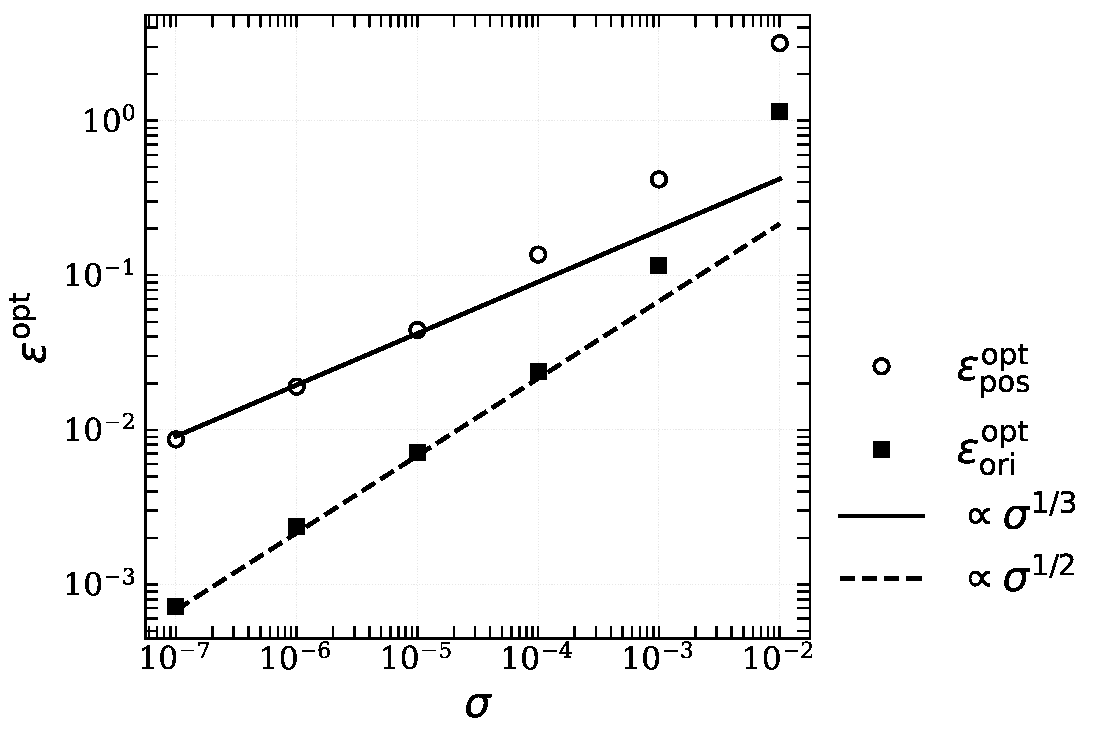}
\caption{Minimum error $\varepsilon^{\mathrm{opt}}$ for each noise intensity $\sigma=10^{-m}\,(m=2,\dots,7)$, minimized over $\Delta r$ and $\Delta t$; circles: position-based method, squares: orientation-based method. Solid and dotted lines show the expected power laws $\varepsilon_{\mathrm{pos}}^{\mathrm{opt}}\propto\sigma^{1/3}$ and $\varepsilon_{\mathrm{ori}}^{\mathrm{opt}}\propto\sigma^{1/2}$.}
\label{fig:sigma}
\end{figure}

To compare the two methods under their best conditions, we define the
minimum error over the time and spatial intervals,
\begin{equation}
  \varepsilon^{\mathrm{opt}} := \min_{\Delta r,\,\Delta t}\varepsilon,
  \label{eq:epsopt}
\end{equation}
where $\varepsilon$ is the relative error in Eq.~\eqref{eq:eps}, and
$\varepsilon_{\mathrm{pos}}^{\mathrm{opt}}$ and
$\varepsilon_{\mathrm{ori}}^{\mathrm{opt}}$ denote its values for the two
methods. Its dependence on the noise intensity is predicted by minimizing
the estimates~\eqref{eq:epsptv} and~\eqref{eq:epsflake}.
Since $\overline{\Delta l}\simeq \|\bm u\|\,\Delta t$, the right-hand sides of
both estimates can be written in the form
\begin{equation}
  \alpha\,\Delta r + \beta\,\Delta t + \gamma\,\frac{\sigma}{D},
  \label{eq:eps_model}
\end{equation}
where $\alpha = \|\nabla G\|/\|G\|$,
$\beta = \bigl(\|G\|^2 + \|\partial G/\partial t\| + \|\nabla G\|\,\|\bm u\|\bigr)/\|G\|$,
$\gamma = 1/\|G\|$, and $D=\Delta r\,\Delta t$ for the position-based method and
$D=\Delta t$ for the orientation-based method.
For the position-based method, the conditions
$\partial\varepsilon_{\mathrm{pos}}/\partial\Delta r
 = \alpha - \gamma\sigma_r/(\Delta r^2\Delta t) = 0$ and
$\partial\varepsilon_{\mathrm{pos}}/\partial\Delta t
 = \beta - \gamma\sigma_r/(\Delta r\,\Delta t^2) = 0$
imply that the three terms in Eq.~\eqref{eq:eps_model} are equal at the
optimum, so that its minimum is
$3\,(\alpha\beta\gamma)^{1/3}\,\sigma_r^{1/3}$.
For the orientation-based method, the noise term does not depend on
$\Delta r$, and the optimization with respect to $\Delta t$ alone gives
$2\,(\beta\gamma)^{1/2}\,\sigma_s^{1/2}+\alpha\,\Delta r$, where the last
term decreases monotonically as $\Delta r$ decreases and is limited only by experimental constraints such as particle size and optical resolution. We therefore expect
$\varepsilon_{\mathrm{pos}}^{\mathrm{opt}}\propto\sigma_r^{1/3}$ and
$\varepsilon_{\mathrm{ori}}^{\mathrm{opt}}\propto\sigma_s^{1/2}$.
Figure~\ref{fig:sigma} shows $\varepsilon^{\mathrm{opt}}$ for
$\sigma=10^{-m}\,(m=2,\dots,7)$. The numerical results follow these laws at
small noise.
At large noise the optimal intervals grow and higher-order truncation terms cause deviations. Note that $\sigma_r$ and $\sigma_s$ are physically distinct, so these results do not directly compare the accuracy of the two methods and a physical comparison is given in section~\ref{subsec:vor_strain}.\\

\subsection{Separation between vorticity and strain accuracy}\label{subsec:vor_strain}

In the following, we compare the two methods for $\sigma_r = 10^{-4}$ and
$\sigma_s = 10^{-3}$, where lengths are nondimensionalized by $1/k$, the
inverse wavenumber of the ABC flow. Although $\sigma_s$ is numerically larger,
the two values are not directly comparable, since $\sigma_r$ is a length and
$\sigma_s$ is an angle. They can, however, be related through a rod of length
$L$ whose ends are located with the same accuracy $\sigma_r$ as the tracer
particles: such a rod yields an orientation noise of
$\sigma_s \simeq \sqrt{2}\,\sigma_r/L$, since the angular error results from
the difference of two independent position errors. The present choice
therefore corresponds to $L \simeq 0.14$, i.e. $L/\sigma_r \simeq 1.4\times10^{3}$,
so that the two values represent equivalent detection accuracy.
The centroids of tracer particle images can be determined accurately~\cite{Tak99, Yam01}.
Assuming a sub-pixel accuracy of 0.1 pixel for particle images
2--5 pixels in diameter, this ratio requires the rod to span approximately
140 pixels in the image.
Two independent ratios should be distinguished here: $kL$, which is set by
the physical sizes of the rod and the flow structures, and $L/\sigma_r$,
which controls the noise. Increasing the optical magnification raises
$L/\sigma_r$ without changing $kL$, at the expense of a narrower field of
view, a shallower depth of field and a shorter tracking time.

\begin{figure*}[t]
  \centering
  \begin{overpic}[width=\linewidth]{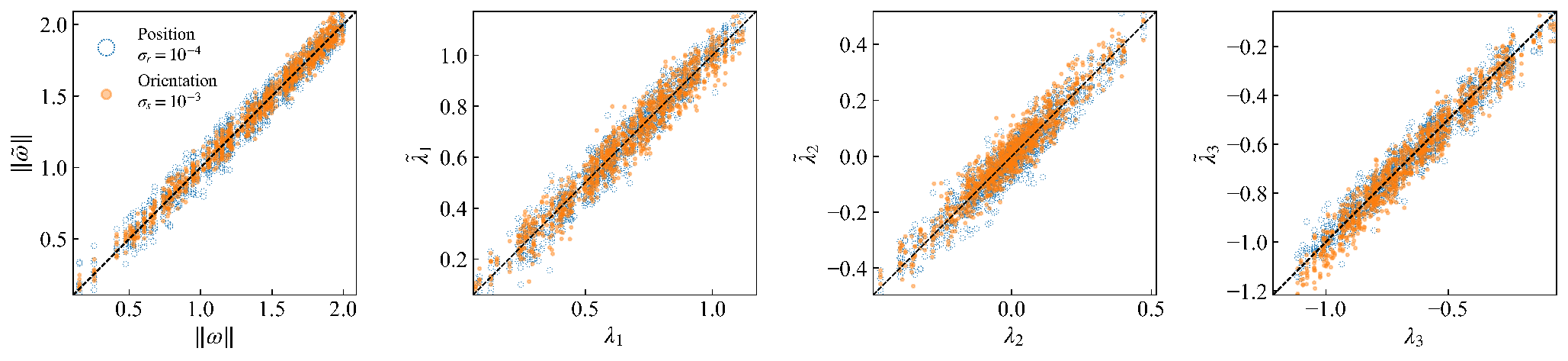}
    \put(0,20){($a$)}
    \put(25,20){($b$)}
    \put(50,20){($c$)}
    \put(75,20){($d$)}
  \end{overpic}
  \caption{Comparison between the true and reconstructed quantities obtained by the position-based method ($\sigma_r=10^{-4}$) and the orientation-based method ($\sigma_s=10^{-3}$).
  ($a$) Vorticity magnitude, $\|\bm{\omega}\|$ versus $\|\tilde{\bm{\omega}}\|$.
  ($b$)--($d$) Eigenvalues of the strain-rate tensor, $\lambda_i$ versus $\tilde{\lambda}_i$ for $i=1$, $2$, and $3$, respectively.
  The dashed lines indicate perfect agreement.}
  \label{fig:vor_strain}
\end{figure*}

Figure~\ref{fig:vor_strain} compares the reconstructed quantities with their true values for the position-based method ($\sigma_r=10^{-4}$) and the orientation-based method ($\sigma_s=10^{-3}$).
Panel ($a$) shows the vorticity magnitude, and panels ($b$)--($d$) show the three eigenvalues of the strain-rate tensor.
Here, $\bm{\omega}$ is the vorticity obtained from the antisymmetric part of $G$, and $\lambda_1 \geq \lambda_2 \geq \lambda_3$ are the eigenvalues of the strain-rate tensor obtained from its symmetric part; the tilde denotes the corresponding quantities obtained from $\tilde{G}$.
The two methods differ clearly in vorticity but not in strain. The orientation-based method reconstructs the vorticity magnitude more accurately than the position-based method (Fig.~\ref{fig:vor_strain}($a$)), whereas both methods give comparable accuracy for all three eigenvalues (Fig.~\ref{fig:vor_strain}($b$)--($d$)).
The same trend holds for the eigenvectors of the strain-rate tensor, whose directional accuracy is also comparable between the two methods (not shown).

\begin{figure}[h]
  \centering
  \begin{overpic}[width=\linewidth]{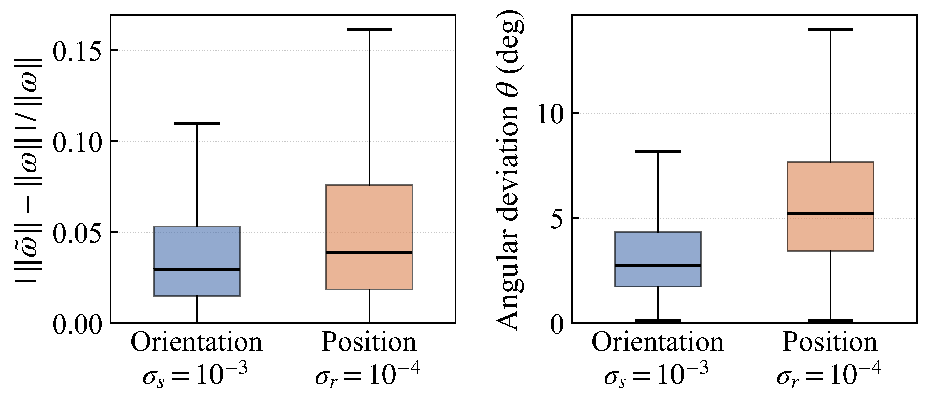}
    \put(0,45){($a$)}
    \put(50,45){($b$)}
  \end{overpic}
  \caption{Accuracy of the reconstructed vorticity obtained by the position-based method ($\sigma_r=10^{-4}$) and the orientation-based method ($\sigma_s=10^{-3}$) for various measurement errors $\sigma$.
  ($a$) Relative error of the vorticity magnitude.
  ($b$) Angular deviation of the vorticity direction, $\theta$, defined in Eq.~\eqref{eq:theta}.}
  \label{fig:vor}
\end{figure}

This advantage in vorticity is confirmed in Fig.~\ref{fig:vor}, which shows the relative error of the vorticity magnitude, $\left|\|\tilde{\bm{\omega}}\|-\|\bm{\omega}\|\right|/\|\bm{\omega}\|$, and the angular deviation of the vorticity direction, $\theta$, defined (in degrees) as
\begin{equation}
  \theta = \frac{180}{\pi}\arccos\left(
  \frac{\bm{\omega}\cdot\tilde{\bm{\omega}}}
       {\|\bm{\omega}\|\,\|\tilde{\bm{\omega}}\|}\right).
  \label{eq:theta}
\end{equation}
Under the conditions ($\sigma_r=10^{-4}$ and $\sigma_s=10^{-3}$), the orientation-based method yields smaller errors than the position-based method in both metrics.
To understand why the advantage appears in vorticity but not in strain, we decompose Eq.~\eqref{eq:orientation} into the vorticity and strain contributions~\citep{Je1922, Br62, Le15}:
\begin{equation}
  \dot{\bm{s}} = \frac{1}{2}\bm{\omega}\times\bm{s}
  - \left[S\cdot\bm{s} - (\bm{s}\cdot S\cdot\bm{s})\,\bm{s}\right],
  \label{eq:jeffery}
\end{equation}
where $S$ is the strain-rate tensor.
Vorticity rotates the particle as a rigid body regardless of its shape,
so the orientation dynamics directly reflects the components of
$\bm{\omega}$ perpendicular to $\bm{s}$, unless strong strain dominates
the rotation and masks the vorticity contribution.
Strain, on the other hand, enters only through the shape-dependent term
(here with the Bretherton parameter $-1$ for thin flakes), which depends
on the differences between the eigenvalues of $S$. The orientation
therefore responds strongly to the dominant strain direction but only
weakly to the others.
Moreover, this term vanishes when $\bm{s}$ is aligned with an
eigenvector of $S$, and strain itself drives the orientation toward
such alignment, so the orientations retain only limited information
about $S$.
In the position-based method, in contrast, both the symmetric and
antisymmetric parts are derived from the same velocity differences,
so no such asymmetry arises.

\subsection{Breakdown of reconstruction with visualization patterns}\label{subsec:SCF}

Flow visualization using thin flake-like particles, such as aluminium flakes or mica, is a classical and convenient technique for qualitative flow observation.
The visualization patterns in this technique emerge because neighbouring particles align their orientations with one another.
From the viewpoint of the present method, however, this alignment is what eliminates the information required for reconstruction. As shown in section~\ref{section:method}, the velocity gradient can be reconstructed only when the particle orientations are sufficiently distinct, and the reconstruction becomes ill-conditioned when they cluster around a common direction.
Moreover, the orientation of each particle is determined by its entire history along its trajectory rather than by the instantaneous flow alone, so the orientation field carries only limited information about the instantaneous velocity gradient. Even whether the orientations converge to an attractor depends on the trajectories~\cite{Sze93p, Sze93, Sze94, Arai25, Arai26}.
In the following, we demonstrate this relationship between the classical visualization and the present reconstruction.

\begin{figure}[h]
\centering
\includegraphics[width=1\linewidth]{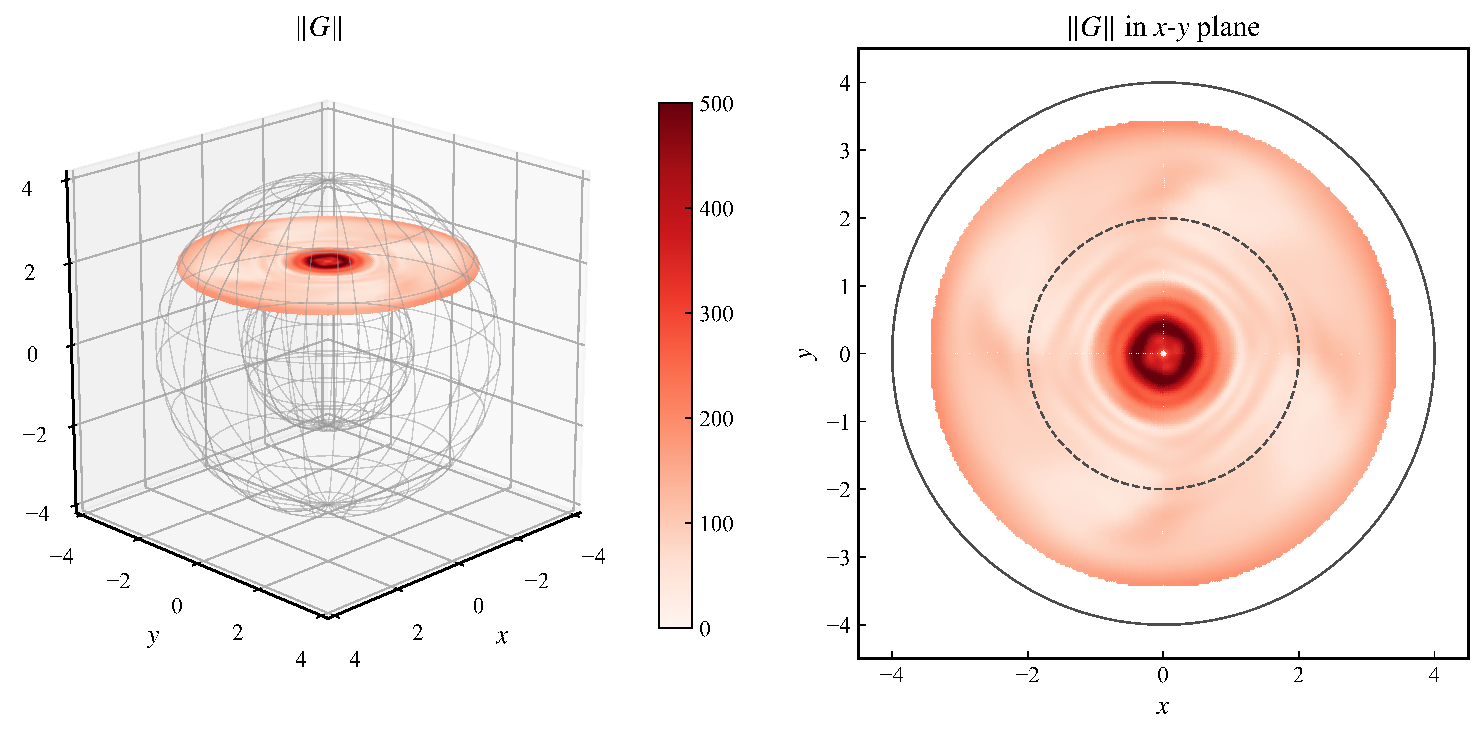}
\caption{Contour map of $\|G\|$ in the plane tangent to the pole of the inner sphere for $(\mathrm{Re},\eta)=(600,1/2)$.}
\label{fig:G_distribution}
\end{figure}

As an application to a real flow in which the orientations degenerate, we consider the wide-gap spherical Couette flow (SCF)~\cite{Egb95}.
Unlike the isotropic orientation distribution in Section~\ref{subsec:ABC}, the tracer orientations in this flow are biased by the flow itself. SCF, the simplest double-shell model relevant to celestial flows, exhibits a variety of states \citep{Wul99, Abb18}. We take the outer sphere ($R_{\mathrm{out}}=4$) at rest and the inner sphere ($R_{\mathrm{in}}=2$) rotating at constant $\Omega$, characterized by $\eta=R_{\mathrm{in}}/R_{\mathrm{out}}$ and $\mathrm{Re}=R_{\mathrm{in}}^2\Omega/\nu$, and we present results for $(\mathrm{Re},\eta)=(600,1/2)$.
At $\mathrm{Re}=600$, the flow has already undergone the first transition from the axisymmetric base flow, and the secondary flow exhibits a characteristic spatial symmetry.
We reconstruct the in-plane velocity gradient tensor corresponding to the visualization plane illuminated by a laser sheet.
Figure~\ref{fig:G_distribution} shows the contour map of $\|G\|$  in the plane tangent to the pole of the inner sphere for the secondary flow.
The flow field was computed as in Ref.~\cite{Go21}, and tracer orientations were initialized as uniform random unit vectors at $t=-T$ and advected from $t=-T$ to $t=0$ following Ref.~\cite{Got11}. 

\begin{figure*}[t]
\centering
\begin{overpic}[width=1\linewidth]{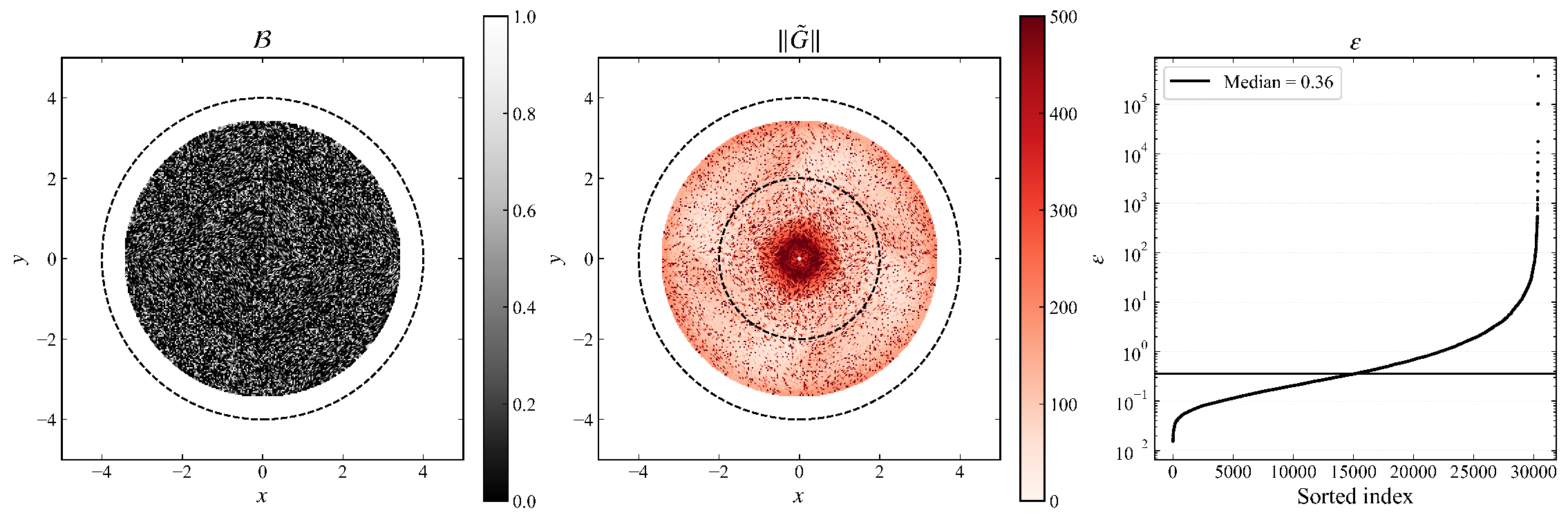}
\put(0,30){(a)-$1$}
\put(35,30){(a)-$2$}
\put(70,30){(a)-$3$}
\end{overpic}\\[2pt]
\begin{overpic}[width=1\linewidth]{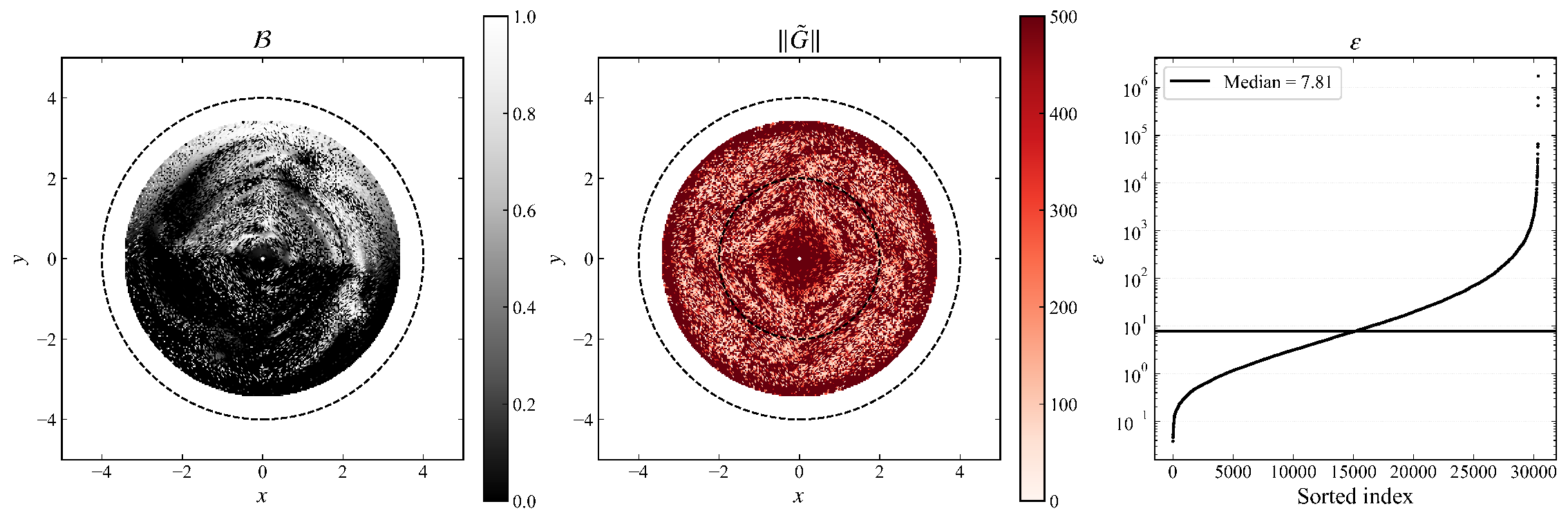}
\put(0,30){(b)-$1$}
\put(35,30){(b)-$2$}
\put(70,30){(b)-$3$}
\end{overpic}

\caption{Comparison between short ($T=\Delta t$; (a)-1--(a)-3) and long
($T=2^{-4}$; (b)-1--(b)-3) advection.
Each particle is plotted at its arrival point and shaded by
(a)-1, (b)-1 the inner product $\mathcal B = \bm R \cdot \bm C = 2 s_y s_z$
for $(\bm C, \bm I) = (\bm e_z, -\bm e_y)$;
(a)-2, (b)-2 the reconstructed $\|\tilde{G}\|$; and
(a)-3, (b)-3 the relative error $\varepsilon$.}
\label{fig:I-G}
\end{figure*}

On this plane we set a $100\times100$ orthogonal grid over a square of side $\sqrt{12}$ ($\Delta r=\sqrt{12}/100\approx3.5\times10^{-2}$), and reconstruct the velocity gradient tensor by the orientation-based method at the grid points inside the outer sphere, using the particles arriving there. The four particles used are the one at the reference point and its three nearest neighbors (at distances $\approx \Delta r$). The results in one quadrant are replicated over the plane by the fourfold azimuthal symmetry. The time interval was $\Delta t=2^{-11}\approx4.9\times10^{-4}$. This corresponds to a time $\Delta t/T_{\mathrm{in}}=O(10^{-2})$ relative to the rotation period $T_{\mathrm{in}}=2\pi/\Omega_{\mathrm{in}}$ of the inner sphere.
Figure~\ref{fig:I-G} contrasts a short advection ((a)-$1$,(a)-$2$,(a)-$3$; $T=\Delta t$) with a long one ((b)-$1$,(b)-$2$,(b)-$3$; $T=2^{-4}$). In panels ((a)-$1$,(b)-$1$), each particle is plotted at its arrival point
and shaded by the inner product $\mathcal B = \bm R \cdot \bm C$, where $\bm C$ is
the unit vector along the optical axis of the camera and $\bm R$ is the
unit vector in the direction of the light reflected by the tracer.
The reflected direction is given by
$\bm R := \bm I - 2(\bm I \cdot \bm s)\bm s$, where $\bm I$ is the unit
vector in the direction of the incident light.
For $(\bm C, \bm I) = (\bm e_z, -\bm e_y)$, this reduces to
$\mathcal B = 2 s_y s_z$.
The spatial distribution of this quantity indicates the spatial bias of the particle orientations and
corresponds to the light-and-dark pattern observed when a laser sheet
illuminates the plane tangent to the inner pole and the reflected light
is recorded from the polar direction.
Panels ((a)-$2$,(b)-$2$) show each particle shaded by the reconstructed $\|\tilde{G}\|$ and panels ((a)-$3$,(b)-$3$) show the relative error $\varepsilon$. For the short exposure the orientations remain diverse, so the pattern of $\mathcal B$ is disordered ((a)-$1$) and the reconstruction is accurate ((a)-$2$ and (a)-$3$). For the long exposure the orientations become locally aligned, so the pattern of $\mathcal B$ traces the flow structure well ((b)-$1$). 
However, this alignment makes neighbouring orientations nearly parallel, driving the coefficient matrix $\mathbf{M}$ toward degeneracy, so that the reconstruction breaks down ((b)-$2$ and (b)-$3$). A characteristic light-and-dark pattern is the visible counterpart of the alignment under which reconstruction breaks down. In this sense, the classical flake visualization and the present quantitative reconstruction are complementary. A clear characteristic pattern marks where the reconstruction fails.

\section{Conclusions and outlook}\label{section:conclusion}

We have presented a method that reconstructs the velocity gradient tensor from non-spherical tracers. Based on the Jeffery-type equation, we have shown that, for incompressible flow, all components of the velocity gradient tensor are uniquely determined from the orientations of four tracers and their time derivatives in the neighbourhood of a single point.

The present method has four notable features. First, it does not use the positions of the tracers and is therefore insensitive to errors in them. Consequently, all components of the velocity gradient tensor can be obtained without measuring orientations throughout a three-dimensional region. It suffices to measure the tracers in the neighbourhood of a reference point.

Second, because the method involves no spatial differencing, measurement noise is amplified by a factor inversely proportional to the temporal interval alone. In the positional PTV-plus-differencing approach, the factor is inversely proportional to the product of the spatial and temporal intervals. As a result, the minimum achievable error scales as $\sigma^{1/2}$ with the noise level $\sigma$ in the present orientation-based method, compared with $\sigma^{1/3}$ for the position-based method, which is advantageous when the noise is small.

Third, the reconstruction accuracy differs between the vorticity and the strain rate only in the orientation-based method, reflecting the structure of Jeffery's equation. The magnitude and direction of the vorticity are reconstructed more accurately than by the position-based method, whereas the strain rate is reconstructed with comparable accuracy.

Finally, the classical  visualization indicates where the velocity gradient is unreconstructable. Using spherical Couette flow, in which the tracer orientations become degenerate, we demonstrated the reconstruction of the velocity gradient tensor. We showed that the emergence of a characteristic light-and-dark pattern is itself a signature of orientation degeneracy: the reconstruction fails where such a pattern appears and succeeds where it does not. The familiar visualization pattern thus serves as a visual indicator of the reconstructability, which is quantified by the physical reconstruction dimension (PRD) \citep{Arai_arxiv}.

The main challenge lies in measuring the orientations of individual tracers. For rod-shaped particles, three-dimensional orientations can already be measured by stereoscopic particle tracking \citep{Ni15}, making them a natural first candidate for experimental application. For flakes, quantitative measurement of each orientation is in principle possible from reflected light recorded from multiple directions, but establishing such a technique remains for future work. In addition, flows with orientational degeneracy require special treatment.

Although independent noise was added at each measurement point, the reconstructions presented here are essentially inverse-crime tests: the synthetic data are generated by the same Jeffery dynamics used for the inversion, so model errors are not assessed. In practice, uncertainty in particle shape and inertial and finite-size effects must also be considered. Since the present method determines the gradient by directly inverting a linear system, it extends naturally to a least-squares formulation when more measurements than unknowns are available, which should improve robustness against such errors. Testing the method on fully resolved simulations of finite-size particles and on experimental data is a natural next step.

\appendix
\section{Error analysis of tracking positions}
\label{section:error_position}

Throughout Appendices~\ref{section:error_position} and~\ref{section:error_orientation}, a tilde denotes a reconstructed quantity, and the error of a quantity $X$ is defined as $\delta X=\tilde X-X$.
Each error is decomposed according to its origin, indicated by a superscript:
``time'' denotes the truncation error of the temporal differencing,
``space'' the error due to the finite spatial separation between particles,
and ``noise'' the error originating from measurement noise.
The forward difference in time along a particle trajectory is written as
\[
\mathcal D_{\Delta t}[X](t)
:=
\frac{X(t+\Delta t)-X(t)}{\Delta t}.
\]

We consider evaluating the velocity gradient tensor based on PTV.
In this method, the velocity is obtained from the particle positions by temporal differencing, and the velocity gradient is then obtained by spatial differencing of the velocity.
The measurement noise of the positions therefore propagates to the velocity gradient through the two stages of velocity evaluation and spatial differencing.

The measured particle position is assumed to contain measurement noise $\bm\epsilon_r$:
\[
\tilde{\bm r}
:=
\bm r
+
\bm\epsilon_r .
\]
The velocity is reconstructed by the forward difference $\tilde{\bm u}=\mathcal D_{\Delta t}[\tilde{\bm r}]$, and its error is decomposed as
\[
\delta\bm u
:=
\tilde{\bm u}
-
\bm u
=
\delta\bm u^{\mathrm{time}}
+
\delta\bm u^{\mathrm{noise}},
\]
where
\begin{equation}
\delta\bm u^{\mathrm{time}}
:=
\mathcal D_{\Delta t}[\bm r]
-
\bm u
\label{eq:deltauD-1}
\end{equation}
is the truncation error of the temporal differencing, and
\[
\delta\bm u^{\mathrm{noise}}
:=
\mathcal D_{\Delta t}[\bm\epsilon_r]
\]
is the error originating from the position noise.
Taylor expansion of the forward difference in Eq.~\eqref{eq:deltauD-1} gives
\begin{equation}
\delta\bm u^{\mathrm{time}}
=
\frac12\,\bm a\,\Delta t
+
O(\Delta t^2),
\label{eq:deltauD-2}
\end{equation}
where $\bm a=\ddot{\bm r}$ is the particle acceleration.
Here, $\bm a$ is regarded as the Eulerian field $\bm a=\partial\bm u/\partial t+(\bm u\cdot\nabla)\bm u$ evaluated at the particle position.

Subsequently, the velocity gradient tensor is reconstructed by componentwise forward differences of the reconstructed velocity,
\[
\tilde G_{ij}
:=
\frac{
\tilde u_i(\bm r+\Delta r_j\bm e_j)
-
\tilde u_i(\bm r)
}
{\Delta r_j},
\]
where $\bm e_j$ is the unit vector in the $x_j$ direction, $\Delta r_j$ is the separation between the neighbouring particles in that direction, and no summation over $j$ is taken.
The error $\delta G:=\tilde G-G$ is decomposed into three components according to their origins:
\begin{equation}
\delta G
=
\delta G^{\mathrm{space}}
+
\delta G^{\mathrm{time}}
+
\delta G^{\mathrm{noise}}.
\label{eq:deltaA}
\end{equation}

The first term of Eq.~\eqref{eq:deltaA} is the truncation error of the spatial differencing itself,
\begin{equation}
\begin{split}
\delta G^{\mathrm{space}}_{ij}
&:=
\frac{u_i(\bm r+\Delta r_j\bm e_j)-u_i(\bm r)}{\Delta r_j}
-
G_{ij}
\\
&=
\frac12\,
\frac{\partial G_{ij}}{\partial x_j}\,
\Delta r_j
+
O(\|\Delta\bm r\|^2),
\end{split}
\label{eq:deltaAD}
\end{equation}
where no summation over $j$ is taken. The norm $\|\cdot\|$ denotes the Euclidean norm for vectors.

The second term of Eq.~\eqref{eq:deltaA} is the temporal truncation error $\delta\bm u^{\mathrm{time}}$, generated at the velocity-evaluation stage, propagated through the spatial differencing:
\[
\begin{split}
\delta G^{\mathrm{time}}_{ij}
&:=
\frac{
\delta u^{\mathrm{time}}_i(\bm r+\Delta r_j\bm e_j)
-
\delta u^{\mathrm{time}}_i(\bm r)
}
{\Delta r_j}
\\
&=
\frac{\partial\,\delta u^{\mathrm{time}}_i}{\partial x_j}
+
O(\|\Delta\bm r\|\,\Delta t).
\end{split}
\]
Substituting Eq.~\eqref{eq:deltauD-2} yields
\[
\delta G^{\mathrm{time}}_{ij}
=
\frac12\,
\frac{\partial a_i}{\partial x_j}\,
\Delta t
+
O(\Delta t^2)
+
O(\|\Delta\bm r\|\,\Delta t),
\]
and, since
\[
\frac{\partial a_i}{\partial x_j}
=
\frac{\partial G_{ij}}{\partial t}
+
(G^2)_{ij}
+
\sum_k u_k\frac{\partial G_{ij}}{\partial x_k},
\]
this becomes
\[
\begin{split}
\delta G^{\mathrm{time}}_{ij}
={}&
\frac12
\biggl(
\frac{\partial G_{ij}}{\partial t}
+
(G^2)_{ij}
+
\sum_k u_k\frac{\partial G_{ij}}{\partial x_k}
\biggr)
\Delta t
\\
&+
O(\Delta t^2)
+
O(\|\Delta\bm r\|\,\Delta t).
\end{split}
\]
Furthermore, since the particle moves by
$\Delta\bm l:=\bm u\,\Delta t+O(\Delta t^2)$
during the time $\Delta t$, we obtain
\begin{equation}
\begin{split}
\delta G^{\mathrm{time}}_{ij}
={}&
\frac12
\biggl(
\frac{\partial G_{ij}}{\partial t}
+
(G^2)_{ij}
\biggr)
\Delta t
+
\frac12
\sum_k
\frac{\partial G_{ij}}{\partial x_k}\,
\Delta l_k
\\
&+
O(\Delta t^2)
+
O(\|\Delta\bm r\|\,\Delta t).
\end{split}
\label{eq:deltaADD}
\end{equation}

The third term of Eq.~\eqref{eq:deltaA} is the position noise amplified through the two stages of temporal and spatial differencing,
\[
\delta G^{\mathrm{noise}}_{ij}
:=
\frac{
\delta u^{\mathrm{noise}}_i(\bm r+\Delta r_j\bm e_j)
-
\delta u^{\mathrm{noise}}_i(\bm r)
}
{\Delta r_j}.
\]
Since the noise at different particles and at different instants is uncorrelated, no cancellation occurs, and
\begin{equation}
\delta G^{\mathrm{noise}}
=
O\!\left(
\frac{\|\bm\epsilon_r\|}
{\|\Delta\bm r\|\,\Delta t}
\right).
\label{eq:deltaAOD}
\end{equation}
The appearance of both $\|\Delta\bm r\|$ and $\Delta t$ in the denominator constitutes the essential difference from the orientation-based method described in Appendix~\ref{section:error_orientation}.

Substituting Eqs.~\eqref{eq:deltaAD}, \eqref{eq:deltaADD} and~\eqref{eq:deltaAOD} into Eq.~\eqref{eq:deltaA}, the error of each component of the velocity gradient tensor is
\begin{equation}
\begin{split}
\delta G_{ij}
={}&
\frac12
\sum_k
\frac{\partial G_{ij}}{\partial x_k}\,
\Delta l_k
+
\frac12
\biggl(
\frac{\partial G_{ij}}{\partial t}
+
(G^2)_{ij}
\biggr)
\Delta t
\\
&+
\frac12
\frac{\partial G_{ij}}{\partial x_j}\,
\Delta r_j
+
O\!\left(
\frac{\|\bm\epsilon_r\|}
{\|\Delta\bm r\|\,\Delta t}
\right)
\\
&+
O\bigl(\|\Delta\bm r\|^2,\ \Delta t^2,\ \|\Delta\bm r\|\,\Delta t\bigr),
\end{split}
\end{equation}
where no summation over $j$ is taken in the first term.
In terms of norms, Eq.~\eqref{eq:PTVerror} reads
\begin{equation}
\begin{split}
\|\delta G\|
={}&
O\!\left(
\Bigl(
\Bigl\|\frac{\partial G}{\partial t}\Bigr\|
+
\|G^2\|
\Bigr)
\Delta t
\right)
+
O\bigl(\|\nabla G\|\,\|\Delta\bm l\|\bigr)
\\
&+
O\bigl(\|\nabla G\|\,\|\Delta\bm r\|\bigr)
+
O\!\left(
\frac{\|\bm\epsilon_r\|}
{\|\Delta\bm r\|\,\Delta t}
\right),
\end{split}
\label{eq:PTVerror_norm}
\end{equation}
which is used for comparison with the orientation-based method.
The norm $\|\cdot\|$ denotes the Frobenius norm for tensors.

\section{Error analysis of tracking orientations}
\label{section:error_orientation}

In the following, we first evaluate the error contained in the time derivative of a single orientation, then add the error introduced when the four orientations are taken from neighbouring particles, and finally show how these errors propagate to the velocity gradient through the linear system $\mathbf M\mathbf x=\mathbf q$ of Eq.~\eqref{eq:reconstruction}.

We write the evolution equation of the unit orientation vector, Eq.~\eqref{eq:orientation}, as
\[
\dot{\bm s}
:=
\bm f(\bm s;G),
\qquad
\lambda:=\bm s\cdot G\cdot\bm s,
\]
where $\bm f$ is linear in $G$.

The four orientations $\bm s^{(i)}$ $(i=1,\dots,4)$ are taken from particles located at $\bm r^{(i)}:=\bm r+\Delta\bm r^{(i)}$, and each measured orientation contains noise $\bm\epsilon_{s}^{(i)}$:
\[
\tilde{\bm s}^{(i)}
:=
\bm s^{(i)}
+
\bm\epsilon_{s}^{(i)}.
\]
From the measured orientations, $\tilde{\mathbf M}$ is constructed from $\tilde{\bm s}^{(i)}$, and $\tilde{\mathbf q}$ is constructed from the forward difference
\[
\tilde{\dot{\bm s}}^{(i)}
:=
\mathcal D_{\Delta t}[\tilde{\bm s}^{(i)}].
\]
The velocity gradient is then obtained from $\tilde{\mathbf M}\tilde{\mathbf x}=\tilde{\mathbf q}$.

The key observation is as follows.
We define
\[
\dot{\tilde{\bm s}}^{(i)}
:=
\bm f\bigl(\tilde{\bm s}^{(i)};G(\bm r)\bigr),
\]
i.e.\ the time derivative that the measured orientation $\tilde{\bm s}^{(i)}$ would have if it evolved under the true velocity gradient $G(\bm r)$ at the reference point.
If the right-hand side of Eq.~\eqref{eq:reconstruction} were constructed from $\dot{\tilde{\bm s}}^{(i)}$, the linear system with the matrix $\tilde{\mathbf M}$ would be satisfied exactly by the true velocity gradient.
Therefore, the noise contained in $\tilde{\mathbf M}$ need not be treated separately, and all errors are contained in the difference
\[
\delta\dot{\bm s}^{(i)}
:=
\tilde{\dot{\bm s}}^{(i)}
-
\dot{\tilde{\bm s}}^{(i)}.
\]
Using the true time derivative $\dot{\bm s}^{(i)}=\bm f(\bm s^{(i)};G(\bm r^{(i)}))$, this difference is decomposed into three components,
\[
\delta\dot{\bm s}^{(i)}
=
\delta\dot{\bm s}^{\mathrm{time}}_{(i)}
+
\delta\dot{\bm s}^{\mathrm{noise}}_{(i)}
+
\delta\dot{\bm s}^{\mathrm{space}}_{(i)},
\]
where
\begin{align*}
\delta\dot{\bm s}^{\mathrm{time}}_{(i)}
:={}&
\mathcal D_{\Delta t}[\bm s^{(i)}]
-
\bm f\bigl(\bm s^{(i)};G(\bm r^{(i)})\bigr),
\\
\delta\dot{\bm s}^{\mathrm{noise}}_{(i)}
:={}&
\mathcal D_{\Delta t}[\bm\epsilon_{s}^{(i)}]
-
\bm f\bigl(\tilde{\bm s}^{(i)};G(\bm r^{(i)})\bigr)
+
\bm f\bigl(\bm s^{(i)};G(\bm r^{(i)})\bigr),
\\
\delta\dot{\bm s}^{\mathrm{space}}_{(i)}
:={}&
\bm f\bigl(\tilde{\bm s}^{(i)};G(\bm r^{(i)})\bigr)
-
\bm f\bigl(\tilde{\bm s}^{(i)};G(\bm r)\bigr).
\end{align*}
The subscript $(i)$ is omitted below when no confusion arises.

\paragraph{Temporal truncation error.}
Taylor expansion of the forward difference gives
\begin{equation}
\delta\dot{\bm s}^{\mathrm{time}}
=
\frac12\,
\ddot{\bm s}\,
\Delta t
+
O(\Delta t^2).
\label{eq:deltadotsD-2}
\end{equation}
Differentiating Eq.~\eqref{eq:orientation}, $\dot{\bm s}=\bm f(\bm s;G)$, with respect to time along the particle trajectory yields
\begin{equation}
\ddot{\bm s}
=
\frac{\partial\bm f}{\partial\bm s}\cdot\dot{\bm s}
+
\bm f\!\left(\bm s;\frac{\mathrm DG}{\mathrm Dt}\right),
\label{eq:ddots}
\end{equation}
where
\[
\frac{\mathrm DG}{\mathrm Dt}
:=
\frac{\partial G}{\partial t}
+
(\bm u\cdot\nabla)G.
\]
The first term of Eq.~\eqref{eq:ddots} is quadratic in $G$ and consists of terms such as $\lambda^2\bm s$, $\lambda\,G\cdot\bm s$ and $G^2\cdot\bm s$.
Since $\|\bm s\|=1$, the Cauchy--Schwarz inequality gives
\begin{equation}
\begin{split}
|\lambda|
=
|\bm s\cdot(G\cdot\bm s)|
&\le
\|\bm s\|\,\|G\cdot\bm s\|
\\
&\le
\|G\|_o
\\
&\le
\|G\|,
\end{split}
\label{eq:lambda_bound}
\end{equation}
where $\|G\|_o:=\max_{\|\bm v\|=1}\|G\cdot\bm v\|$ is the operator norm, which is bounded by the Frobenius norm $\|G\|$.
Hence $\lambda=O(\|G\|)$, and consequently $\bm f=O(\|G\|)$ and $\partial\bm f/\partial\bm s=O(\|G\|)$.
Hence
\[
\ddot{\bm s}
=
O(\|G\|^2)
+
O\!\left(\Bigl\|\frac{\partial G}{\partial t}\Bigr\|\right)
+
O\bigl(\|(\bm u\cdot\nabla)G\|\bigr).
\]
Using $\Delta\bm l=\bm u\,\Delta t+O(\Delta t^2)$, Eq.~\eqref{eq:deltadotsD-2} becomes
\begin{equation}
\begin{split}
\delta\dot{\bm s}^{\mathrm{time}}
={}&
O\!\left(
\Bigl(
\|G\|^2
+
\Bigl\|\frac{\partial G}{\partial t}\Bigr\|
\Bigr)
\Delta t
\right)
+
O\bigl(\|\nabla G\|\,\|\Delta\bm l\|\bigr).
\end{split}
\label{eq:deltadotsD}
\end{equation}

\paragraph{Measurement noise.}
The second and third terms of $\delta\dot{\bm s}^{\mathrm{noise}}$ arise naturally from comparing $\tilde{\dot{\bm s}}$ with $\dot{\tilde{\bm s}}$, which is evaluated at the noisy orientation.
Linearizing $\bm f$ with respect to $\bm s$ and neglecting terms of second order in $\bm\epsilon_s$, we obtain
\begin{equation}
\delta\dot{\bm s}^{\mathrm{noise}}
=
\frac{\bm\epsilon_s(t+\Delta t)-\bm\epsilon_s(t)}{\Delta t}
+
\bm\Gamma,
\label{eq:deltadotsO}
\end{equation}
where
\[
\bm\Gamma
=
-
\frac{\partial\bm f}{\partial\bm s}\cdot\bm\epsilon_s+O(\|\bm\epsilon_s\|^2).
\]
Since $\partial\bm f/\partial\bm s=O(\|G\|)$, the term $\bm\Gamma$ is of order $\|G\|\,\|\bm\epsilon_s\|$, and
\[
\delta\dot{\bm s}^{\mathrm{noise}}
=
O\!\left(\frac{\|\bm\epsilon_s\|}{\Delta t}\right)
+
O\bigl(\|G\|\,\|\bm\epsilon_s\|\bigr)+O(\|\bm\epsilon_s\|^2).
\]

\paragraph{Spatial separation of neighbouring particles.}
The present method requires four orientations and their time derivatives measured at the same instant under the same velocity gradient $G$, but it is in principle impossible to obtain four independent orientations simultaneously at a single position.
Neighbouring particles are therefore used, and each of them experiences $G(\bm r^{(i)})$ instead of $G(\bm r)$.
Since $\bm f$ is linear in $G$,
\begin{equation}
\begin{split}
\delta\dot{\bm s}^{\mathrm{space}}_{(i)}
&=
\bm f\bigl(\tilde{\bm s}^{(i)};G(\bm r^{(i)})-G(\bm r)\bigr)
\\
&=
\bm f\bigl(\tilde{\bm s}^{(i)};(\Delta\bm r^{(i)}\cdot\nabla)G\bigr)
+
O(\|\Delta\bm r^{(i)}\|^2)
\\
&=
O\bigl(\|\nabla G\|\,\|\Delta\bm r^{(i)}\|\bigr)+O(\|\Delta\bm r^{(i)}\|^2).
\end{split}
\label{eq:deltadotsI}
\end{equation}

\paragraph{Propagation to the velocity gradient.}
Let $\mathbf x$ be the true solution corresponding to $G(\bm r)$, and let $\mathbf q$ be constructed from $\dot{\tilde{\bm s}}^{(i)}$ by Eq.~\eqref{eq:q}, so that $\tilde{\mathbf M}\mathbf x=\mathbf q$ holds exactly.
The error of the right-hand side, $\delta\mathbf q:=\tilde{\mathbf q}-\mathbf q$, is composed of $\delta\dot{\bm s}^{(i)}$ $(i=1,\dots,4)$, so that $\|\delta\mathbf q\|=O(\max_i\|\delta\dot{\bm s}^{(i)}\|)$, and its components are denoted by $\delta\mathbf q^{\mathrm{time}}$, $\delta\mathbf q^{\mathrm{noise}}$ and $\delta\mathbf q^{\mathrm{space}}$ according to their origins.
Subtracting $\tilde{\mathbf M}\mathbf x=\mathbf q$ from $\tilde{\mathbf M}\tilde{\mathbf x}=\tilde{\mathbf q}$ gives
\[
\delta\mathbf x
:=
\tilde{\mathbf x}-\mathbf x
=
\tilde{\mathbf M}^{-1}\delta\mathbf q .
\]
Denoting the largest singular value of $\tilde{\mathbf M}^{-1}$ by $\kappa:=1/\sigma_{\min}(\tilde{\mathbf M})$, the error is bounded, up to a constant factor of order unity relating $\|\delta\mathbf x\|$ and $\|\delta G\|$, by
\begin{equation}
\begin{split}
\|\delta G\|
\le
\kappa
\bigl(
\|\delta\mathbf q^{\mathrm{time}}\|
+
\|\delta\mathbf q^{\mathrm{noise}}\|
+
\|\delta\mathbf q^{\mathrm{space}}\|
\bigr).
\end{split}
\label{eq:deltaAflake}
\end{equation}
The quantity $\kappa$ depends on the geometric configuration of the orientations and increases as the orientations approach degeneracy.
Substituting Eqs.~\eqref{eq:deltadotsD}, \eqref{eq:deltadotsO} and~\eqref{eq:deltadotsI} into Eq.~\eqref{eq:deltaAflake}, the error of the reconstructed velocity gradient tensor is evaluated as
\begin{equation}
\begin{split}
\|\delta G\|
\le
\kappa\,
\biggl[\,
&O\!\left(
\Bigl(
\Bigl\|\frac{\partial G}{\partial t}\Bigr\|
+
\|G\|^2
\Bigr)
\Delta t
\right)
+
O\bigl(\|\nabla G\|\,\|\Delta\bm l\|\bigr)
\\
&+
O\bigl(\|\nabla G\|\,\|\Delta\bm r\|\bigr)
+
O\!\left(\frac{\|\bm\epsilon_s\|}{\Delta t}\right)
+
O\bigl(\|G\|\,\|\bm\epsilon_s\|\bigr)
\,\biggr],
\end{split}
\label{eq:orientationerror}
\end{equation}
where $\|\Delta\bm r\|:=\max_i\|\Delta\bm r^{(i)}\|$.
The contributions of $\lambda^2$ and $\lambda G$ in $\ddot{\bm s}$ are included in the $O(\|G\|^2\Delta t)$ term.
The particle positions are also subject to measurement noise
$\bm\epsilon_r^{(i)}$, so that the actual separation is
$\Delta\bm r^{(i)}+\bm\epsilon_r^{(i)}$.
Compared with Eq.~\eqref{eq:PTVerror_norm}, the noise term is not divided by $\|\Delta\bm r\|$, and $\|\Delta\bm r\|$ appears only in the numerator; hence reducing the particle separation improves the spatial accuracy without amplifying the measurement noise.
In the orientation-based method, however, $\bm\epsilon^{(i)}_r$ is not
amplified, since it does not enter through division by $\Delta\bm r^{(i)}$;
it merely shifts the sampling point, and its contribution is of order
$O(\|\Delta\bm r^{(i)}+\bm\epsilon_r^{(i)}\|)$.
Assuming $\|\bm\epsilon_r^{(i)}\|\lesssim\|\Delta\bm r^{(i)}\|$, this reduces to
$O(\|\Delta\bm r^{(i)}\|)$.

\begin{acknowledgments}
The authors would like to thank Mr. F. Goto and Mr. T. Inagaki for their pilot surveys.
We gratefully acknowledge discussions with Prof. Masako Sugihara-Seki.
This work was supported in part by JSPS KAKENHI Grant No. JP24K07331 and by the Kansai University Grant-in-Aid for Progress of Research in Graduate Course, 2024.
\end{acknowledgments}

\bibliographystyle{ieeetran}
\bibliography{SCFreconst}

\end{document}